\documentclass[journal=jpclcd,manuscript=article]{achemso}
\usepackage[version=3]{mhchem}
\usepackage{graphicx}
\usepackage{bm}
\usepackage{amsmath}
\usepackage{amssymb}
\usepackage{braket}
\usepackage{multirow}
\usepackage{enumerate}
\usepackage{threeparttable}
\usepackage{interval}
\usepackage{xcolor}
\usepackage{xr}
\usepackage{natbib}

\makeatletter
\newcommand*{\addFileDependency}[1]{
  \typeout{(#1)}
  \@addtofilelist{#1}
  \IfFileExists{#1}{}{\typeout{No file #1.}}
}

\newsavebox\myboxA
\newsavebox\myboxB
\newlength\mylenA

\newcommand*\xoverline[2][0.75]{%
    \sbox{\myboxA}{$\m@th#2$}%
    \setbox\myboxB\null
    \ht\myboxB=\ht\myboxA%
    \dp\myboxB=\dp\myboxA%
    \wd\myboxB=#1\wd\myboxA
    \sbox\myboxB{$\m@th\overline{\copy\myboxB}$}
    \setlength\mylenA{\the\wd\myboxA}
    \addtolength\mylenA{-\the\wd\myboxB}%
    \ifdim\wd\myboxB<\wd\myboxA%
       \rlap{\hskip 0.5\mylenA\usebox\myboxB}{\usebox\myboxA}%
    \else
        \hskip -0.5\mylenA\rlap{\usebox\myboxA}{\hskip 0.5\mylenA\usebox\myboxB}%
    \fi}
\makeatother

\newcommand*{\myexternaldocument}[1]{
    \externaldocument{#1}
    \addFileDependency{#1.tex}
    \addFileDependency{#1.aux}
}

\myexternaldocument{SI}
\def\br{{\bf r}}
\def\bR{{\bf R}}

\author{Guorong Weng}
\affiliation[University of California, Santa Barbara]
{Department of Chemistry and Biochemistry, University of California, Santa Barbara, California, 93106-9510, United States}
\author{Amanda Pang}
\affiliation[University of California, Santa Barbara]
{Department of Chemistry and Biochemistry, University of California, Santa Barbara, California, 93106-9510, United States}
\altaffiliation{Current institution: University of Pennsylvania, Philadelphia, 19104, United states}
\author{Vojt\v{e}ch Vl\v{c}ek}
\email{vlcek@ucsb.edu}
\affiliation[University of California, Santa Barbara]
{Department of Chemistry and Biochemistry, University of California, Santa Barbara, California, 93106-9510, United States}
\alsoaffiliation[University of California, Santa Barbara]
{Materials Department, University of California, Santa Barbara, California, 93106-5050, United States}

\title{Spatial Decay and Limits of Quantum Solute-Solvent Interactions}


\begin{document}

\begin{tocentry}
    \centering
    \includegraphics[width=4.5cm]{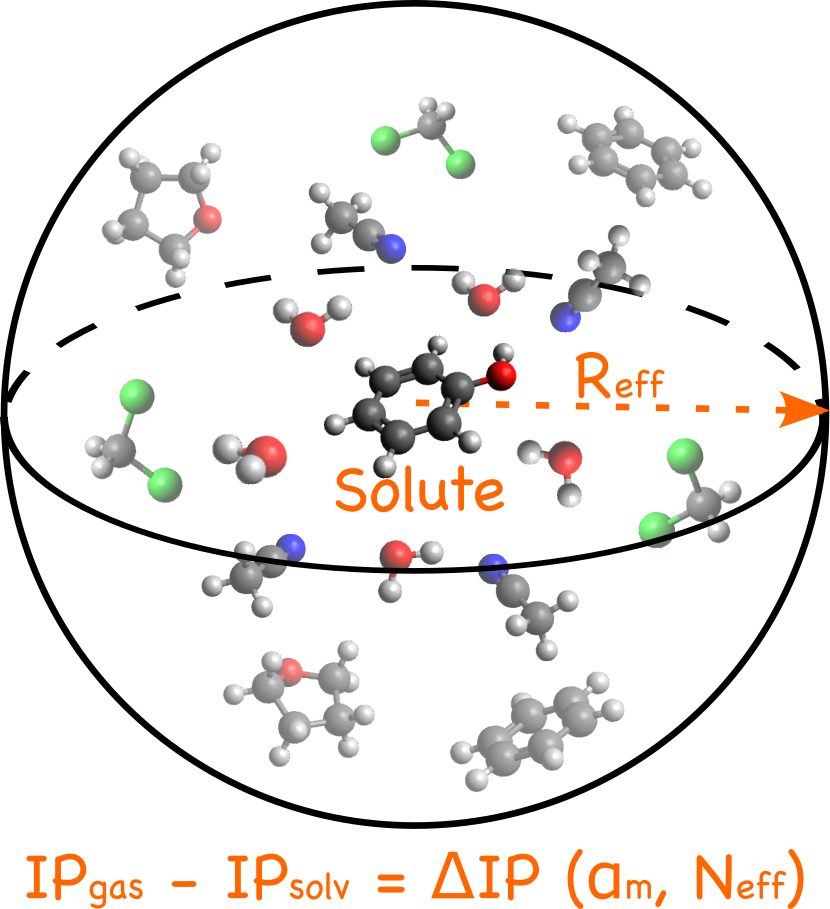}
    \label{fig:TOCentry}
\end{tocentry}

\begin{abstract}
Molecular excitations in the liquid-phase environment are renormalized by the surrounding solvent molecules. Herein, we employ the $GW$ approximation to investigate the solvation effects on the ionization energy of phenol in various solvent environments. The electronic effects differ by up to 0.4 eV among the five investigated solvents. This difference depends on both the macroscopic solvent polarizability and the spatial decay of the solvation effects. The latter is probed by separating the electronic subspace and the $GW$ correlation self-energy into fragments. The fragment correlation energy decays with increasing intermolecular distance and vanishes at $\sim$9 {\AA}, and this pattern is independent of the type of solvent environment. The 9 {\AA} cutoff defines an effective interacting volume within which the ionization energy shift per solvent molecule is proportional to the macroscopic solvent polarizability. Finally, we propose a simple model for computing the ionization energies of molecules in an arbitrary solvent environment.   
\end{abstract}

Liquid phase is the most practical environment for treating multicomponent chemical systems. Applications in synthetic chemistry and material processing rely on the dissolving ability, primarily linked to the solvent's permanent dipole, i.e., the polarity. In the context of electronic excitations (e.g., in spectroscopy) or charge transport, the dynamical electric polarizability represents another fundamental property characterizing the solvent medium; it is related to the dynamical induced dipoles that screen the electron-electron interactions. The charge carriers are ``dressed'' by mutual interactions, forming quasiparticles (QP),\cite{fetter2003,Martin2016} and the electronic structure is renormalized. Specifically, the presence of solvent shifts the excitation energies of the solute molecule and leads to their finite excitation lifetimes. The effects on molecular energy levels of chromophores are commonly known as \textit{solvatochromic shifts}.\cite{Buncel1990,Reichardt1994} In experiments, these shifts can be measured by light absorption and emission spectroscopy as well as photoelectron spectroscopy.

The detailed microscopic understanding of solvatochromic shifts requires efficient and accurate first-principles simulations. State-of-the-art theoretical approaches for simulating solute-solvent systems are mainly formulated within the embedding framework: quantum mechanical approaches, including quantum chemistry methods\cite{Kosenkov2011,Isborn2011,Bistafa2012,Ghosh2012,Caricato2012,Schwabe2012,Bose2016,Sadybekov2017,Ghosh2017,Ren2017,Chakraborty2017,Hrsak2018,Lu2018,Ren2019,Caricato2020,Folkestad2020,Goletto2021} and time-dependent density functional theory (TDDFT),\cite{Santoro2006,Barone2008,Santoro2010,Sok2011,Dargiewicz2012,Improta2012,Isborn2012,Marenich2015,Guido2015,Bi2017,Rubesova2017,Preiss2018,Chakravarty2022} apply only to the ``core region,'' which consists of the solute molecule and sometimes a small number of proximal solvent molecules;\cite{Santoro2006,Kosenkov2011,Ghosh2012,Marenich2015,Bose2016,Ren2017,Ren2019,Folkestad2020,Goletto2021} the solvent environment is treated most often with either the polarizable continuum model\cite{Cramer1999,Tomasi2005,Mennucci2012,Lipparini2016} (PCM) or classical molecular mechanics.\cite{Gordon2012,Pruitt2014} However, high-accuracy quantum chemistry approaches are limited to small molecules due to the steep cost-scaling; TDDFT results from practical implementations strongly depend on the choice of the exchange-correlation functional. For the environment, the use of implicit solvent models (e.g., PCM) does not allow detailed investigations, for instance, about the spatial decay of the solute-solvent interactions. Furthermore, current studies\cite{Kosenkov2011,Isborn2011,Bistafa2012,Caricato2012,Schwabe2012,Ghosh2017,Ren2017,Lu2018,Ren2019,Folkestad2020,Goletto2021,Santoro2006,Santoro2010,Sok2011,Improta2012,Isborn2012,Marenich2015,Guido2015,Bi2017,Preiss2018,Chakravarty2022,Zuehlsdorff2019} focus mainly on the optical absorption and emission processes but do not provide direct information about the absolute energies of electronic levels, i.e., ionization potential (IP) and electron affinity (EA). These levels represent the electron-donating and accepting abilities of the solute molecule. The development of liquid microjet photoelectron spectroscopy\cite{Winter2006,Seidel2011,Tentscher2015,Suzuki2019} allows direct IP measurements for solvated molecules. However, this technique applies mainly to aqueous solutions due to the volatility of most organic solvents. An affordable and accurate theoretical approach is in demand for predicting the IP and EA of molecules in diverse liquid environments.

Many-body Green's function methods\cite{fetter2003,Martin2016} provide direct access to single-quasiparticle energies, provided that the system is weakly or moderately correlated. Indeed, the $GW$ approximation,\cite{Hedin1965,Aryasetiawan1998,Onida2002,Schindlmayr2006,Golze2019} even at the lowest order expansion in which the electronic correlation is described merely through charge density fluctuations, yields IPs that agree with experiments for most molecular systems.\cite{vanSetten2015,Caruso2016,Govoni2018} Recent developments in efficient algorithms\cite{Pham2013,Govoni2015,Bruneval2016,Liu2016,Wilhelm2018,Yang2019,Forster2020,Kim2020,Gao2020,Wilhelm2021,Forster2021,Umari2022} and high performance computing,\cite{DelBen2019,Ben2020,Yu2022} especially the linear-scaling stochastic formalism,\cite{Neuhauser2014_prl,Vlcek2017,Vlcek2018_prb,Vlcek2019,Vlcek2018_prm,Brooks2020,Weng2020,Romanova2020,Weng2021,Romanova2022} have enabled large-scale $GW$ calculations for systems with thousands of electrons. Within the stochastic $GW$ framework,\cite{Neuhauser2014_prl,Vlcek2018_prb,Vlcek2019} our previous work\cite{Weng2021} established an efficient approach for computing the photoemission spectra (i.e., IPs) of various solvated molecules, in which the solute and the solvent environment (containing $\sim$1000 electrons) are treated on the same footing. Excellent agreements with experiments and other comparable methods have been achieved for molecules solvated by water. 

In this work, we investigate the molecular ionization in various polarizable solvents and the spatial decay of the solvation electronic effects on the ionization energy. We use regionally-localized\cite{Weng2022} Pipek-Mezey\cite{Pipek1989,Lehtola2014,Jonsson2017} orbitals to separate the electronic subspaces and decompose the correlation contributions into fragments. The methodology is exemplified on a phenol molecule in five different solvents, with geometric structures generated from molecular dynamics (MD) simulations (details are provided in the supporting information). The energy shifts of the phenol's IP are computed and related to the macroscopic solvent polarizability. The rapid convergence of the IP shifts with respect to the number of surrounding solvent molecules indicates that the solute-solvent interactions vanish at some distance, which is fairly uniform across vastly different types of solvents. From the decay of the $GW$ correlation self-energy, we identify an effective interacting radius for the solute molecule to interact with the induced charge density from the environment. Within the effective interacting volume, we find that the IP shift per solvent molecule is proportional to the polarizability volume calculated for each solvent. Finally, a simple solvation model is proposed for computing the IP of molecules in an arbitrary solvent environment.

In the Green's function formalism,\cite{fetter2003,Martin2016} the electron-electron interactions are represented by the nonlocal and dynamical exchange-correlation self-energy, $\Sigma_{\rm{X}}$ and $\Sigma_{\rm{C}}$. In practice, we compute $\Sigma_{\rm{X}}$ and $\Sigma_{\rm{C}}$ as perturbative corrections to the mean-field eigenvalue yielding the following QP energy
\begin{equation}
   \varepsilon_j^{\rm QP} = \braket{\phi_j |\varepsilon_j^{0} - \hat{v}_{\rm{xc}} + \hat{\Sigma}_{\rm{X}} + \hat{\Sigma}_{\rm{C}} (\omega=\varepsilon_j^{\rm QP}) | \phi_j}
\label{eq:qpe}
\end{equation}
Here, $\phi_j$ is a reconstructed molecular state on the solute,\cite{Weng2021} and $\varepsilon^0_j$ comes from an auxiliary density functional theory\cite{Hohenberg1964, Kohn1965} (DFT) calculation of the isolated solute;\cite{Weng2021} $\hat{v}_{\rm{xc}}$ is the PBE exchange-correlation potential,\cite{Perdew1996} and $\hat{\Sigma}_{\rm{X}}$ is the nonlocal exchange interaction equivalent to the Fock operator;\cite{Martin2016} $\hat{\Sigma}_{\rm{C}}(\omega)$ is the frequency-dependent correlation self-energy. This work employs the one-shot $GW$ ($G_0W_0$) approach, in which the correlation corresponds to the potential due to charge density fluctuations. In the following, we use $\Sigma_{\rm{C}}$ to denote the expectation value $\braket{\phi_j|\hat{\Sigma}_{\rm{C}}|{\phi_j}}$, where the spatial coordinates are integrated out.

We demonstrated the separation of $\Sigma_{\rm{C}}$ into the molecular and the environmental contributions in ref~\citenum{Weng2021}. In this work, we generalize this separation to multifragments in the solute-solvent systems. The definition of a fragment is arbitrary and in the remainder of this text, a fragment refers to a solvation shell consisting of one or multiple solvent molecules. The electronic subspace of a fragment is represented by the following projector
\begin{equation}
   \hat{P}^k = \sum_i^{N_{\rm s}} \ket{\psi_i^{k}} \bra{\psi_i^{k}}
\label{eq:space_frag}
\end{equation}
where $\psi^{k}$ form a localized basis, and $N_s$ includes all the valence electrons of the $k^{\rm{th}}$ fragment. The corresponding electronic subspace is sufficiently defined by the full set of occupied states based on the ``local density fluctuations'' assumption: the perturbed and time-evolved $\psi^k$ stay localized on the $k^{\rm{th}}$ fragment. In other words, we assume no intermolecular charge transfer happens, and all the density fluctuations remain on the fragment. This assumption is reasonable when there is no apparent donor-acceptor character found in a van der Waals bound molecular system. It follows that the time-dependent charge-density fluctuations $\delta n(\br,t)$ are decomposed into fragments
\begin{equation}
\delta n(\br,t) = \sum_k^{N_{\rm frag}} \delta n^k(\br,t)
\label{eq:deltadens}
\end{equation}
where $N_{frag}$ denotes the number of fragments, and $\delta n^k(\br,t)$ is the density fluctuations contributed by the $k^{\rm{th}}$ fragment.

Since the $G_0W_0$ correlation self-energy stems from the charge density-density interactions (i.e., induced dipole interactions), the $\Sigma_{\rm{C}}$ can be immediately written as
\begin{equation}
   \Sigma_{\rm{C}}[\delta n(t),t] = \sum_{k}^{N_{\rm frag}} \Sigma^k_{\rm{C}} [\delta n^k(\br,t), t]
\label{eq:dce_frag}
\end{equation}
In this work, we use the stochastic $GW$ method\cite{Neuhauser2014_prl,Vlcek2018_prb,Vlcek2019} to compute the correlation self-energy contributed from a specific fragment, e.g., a solvation shell at distance $\bR^k$, and then study the decay of $\Sigma_{\rm{C}}^k$ as a function of $\bR^k$.

The construction of $\hat{P}^k$, the decomposition of $\delta n(\br,t)$, and the calculation of $\Sigma^k_{\rm{C}}$ are detailed in the Theory and Methodology section in the Supporting Information.

\begin{figure}
    \centering
    \includegraphics[width=\textwidth]{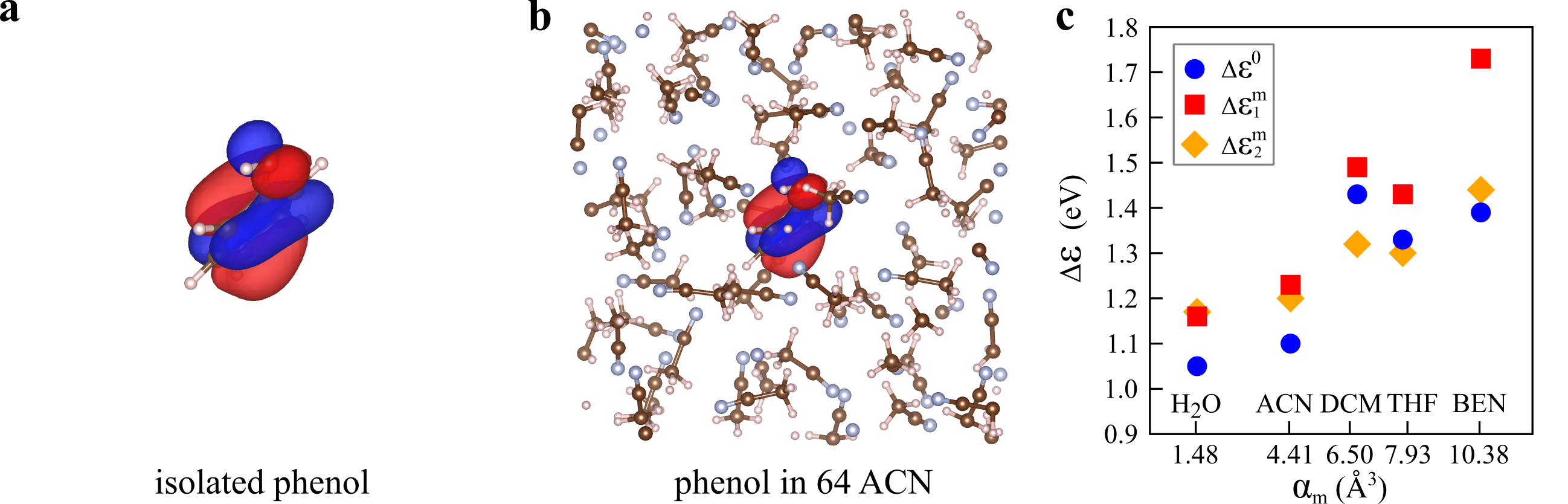}
    \caption{(a) HOMO of the isolated phenol molecule. (b) Reconstructed HOMO of the phenol molecule surrounded by 64 ACN molecules in the simulation cell. (c) IP shifts $\Delta \varepsilon$ plotted as a function of the mean polarizability volume: $\Delta \varepsilon^0$ represents the first-principles results calculated as $\varepsilon^{\rm{solv}}-\varepsilon^{\rm{iso}}$; $\Delta \varepsilon^{\rm{m}}_1$ and $\Delta \varepsilon^{\rm{m}}_2$ are derived from the proposed solvation model, where $\Delta \varepsilon^{\rm{m}}_1$ uses $N_{\rm{eff}}$ derived from the solvent mass density, while $\Delta \varepsilon^{\rm{m}}_2$ uses the average $N_{\rm{eff}}$ over five selected snapshots from the MD trajectory.}
    \label{fig:fig1}
\end{figure}

To practically investigate the energy shifts contributed by the solvent environment, we explore the vertical ionization potential (corresponding to the negative of the HOMO QP energy) of phenol in five different solvents: water (H$_2$O), acetonitrile (ACN), dichloromethane (DCM), tetrahydrofuran (THF), and benzene (BEN). The solute-solvent systems are constructed and propagated using a cubic cell with lateral dimensions of 18$-$21 {\AA} in MD simulations (see Figure~\ref{fig:md_big} and Table~\ref{tab:num_solv}). The solvent environment contributes to the solute's electronic structure in two ways: direct electron-electron interactions and structural relaxations. The latter also involves direct electron/hole-vibrational couplings, which are, however, not considered in this work. The structural effects are thus reduced to conformational changes and the corresponding QP energies of the isolated phenol molecule ($\varepsilon^{\rm{iso}}$), with molecular structures extracted from snapshots (Figure~\ref{fig:md_small} and Figure~\ref{fig:md_big}) of the MD trajectories. Figure~\ref{fig:fig1}a shows the HOMO of an isolated phenol molecule, which is obtained by simply removing all the solvent molecules (Figure~\ref{fig:fig1}b). The $\varepsilon^{\rm{iso}}$ results do not differ too much among the five solvents (see Table~\ref{tab:qpenergy}), i.e., the solvent-induced structural changes of phenol are consistent regardless of the actual chemical environment in the liquid phase. Note that flexible molecules, e.g., phenylalanine with a chain group attached to the aromatic ring, will display a more pronounced dependence on the structural variation (see our reported results in ref~\citenum{Weng2021}). Since we focus on the purely electronic contributions stemming from the dynamical electronic interactions with the solvent environment, phenol thus appears to be an appropriate choice of test systems.

The solvation many-body effects on the IP of phenol correspond to the energy shifts, $\Delta \varepsilon$, defined as the QP energy difference between the solvated and the isolated HOMO ($\varepsilon^{\rm{solv}}- \varepsilon^{\rm{iso}}$). For each solvent environment, we sample $\Delta \varepsilon$ using five snapshots and obtain the average value (see Figure~\ref{fig:IP_big}). The standard deviation of $\Delta \varepsilon$ is lower than that of $\varepsilon^{\rm solv}$, indicating that the sampling of $\Delta \varepsilon$ represents the average solvation effects along the MD trajectory. A significant part of these effects is due to the dynamical screening by the induced charge density fluctuations of the solvent environment.\cite{Weng2021} Hence, it stands to reason that $\Delta \varepsilon$ is related to the macroscopic solvent polarizability, which is further related to the refractive index $n_r$ of the liquid by the Lorentz-Lorenz equation, $\alpha_m = 3({n_r}^2-1)/4\pi({{n_r}^2+2})$. $\alpha_m$ is the mean polarizability volume, and $N_v$ is the number of molecules per unit volume. The derivation of $N_v$ takes the solvent's mass density at 293 K. Both $n_r$ and mass density data are readily accessible in the CRC Handbook of Chemistry and Physics.\cite{crchandbook} In Figure~\ref{fig:fig1}c, the $\Delta \varepsilon^0$ (blue circles, first-principles values) averaged over five snapshots of each solvated system is plotted with respect to the corresponding $\alpha_m$. While there is a general trend, the dependence is not straightforward: most importantly, the DCM environment exhibits the largest $\Delta \varepsilon^0$ (IP shifts), despite its polarizability volume being in the middle of the range. The results in Figure~\ref{fig:fig1}c are then compared with simulations using a smaller cell and fewer solvent molecules (see Figures~\ref{fig:md_small} and~\ref{fig:IP_small}). Even though the numbers of solvent molecules are (nearly) doubled (Table~\ref{tab:num_solv}), the $\Delta \varepsilon^0$ are increased by at most 15\% (see Figure~\ref{fig:cell_compare}). It implies that the solute-solvent many-body interactions are converging rapidly with an increasing number of solvent molecules, at least at the $GW$ level. In fact, we expect that these intermolecular interactions decay with distance, and the decay is governed by the polarizability of the solvent molecules.

Next, we investigate the distance ($\bR^k$)-dependence of the correlation self-energy ($\Sigma^k_{\rm{C}}$) contributed from the liquid environment. At the $G_0W_0$ level, this corresponds to the polarization screening effects stemming from the induced time-dependent dipoles on the solvent molecules. The solvent environment is fragmented into solvation shells (Figure~\ref{fig:shell_all}), each of which is represented by a set of localized orbitals (eq~\eqref{eq:space_frag}). Note that the shell in this work is constructed by a cluster of solvent molecules at a similar distance ($\pm$0.2 \AA). We set integer distances ranging from 3 to 10 {\AA} for the fragment selection, while the actual distance $\bR^k$ is derived by averaging the distances of all molecules within the same shell. In total, six shells are chosen from a typical snapshot of each solvated system, and most of them contain more than one molecule in order to cancel out the solvent orientation effects (Figure~\ref{fig:shell_all}). 
\begin{figure}
    \centering
    \includegraphics[width=.8\textwidth]{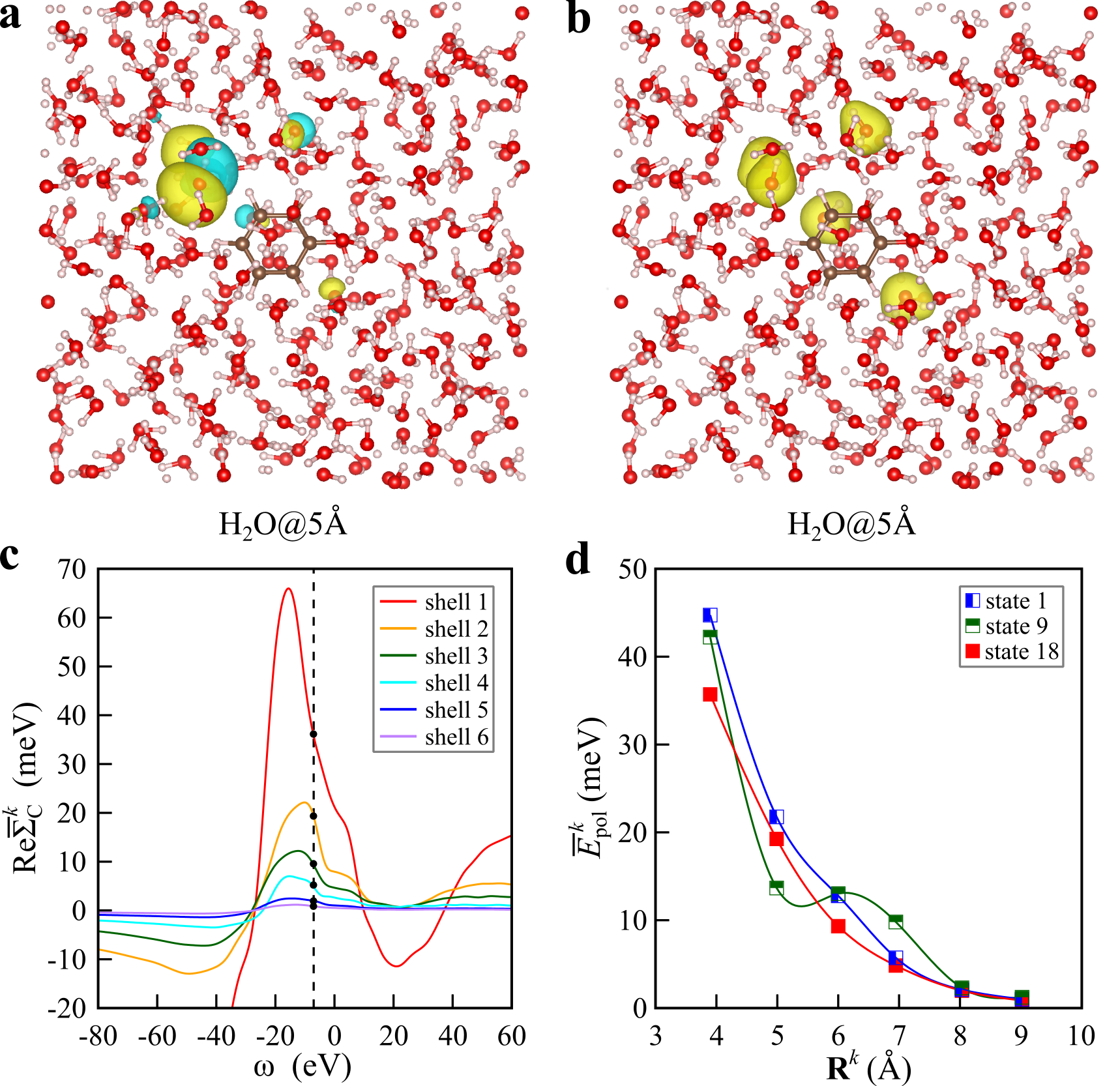}
    \caption{(a) A typical localized orbital on the fragment with 5 H$_2$O molecules at $\sim$5 {\AA}. (b) Localized electron density on the fragment with 5 H$_2$O molecules at $\sim$5 {\AA}. (c) Real part of the normalized fragment correlation self-energy for each H$_2$O shell. The dashed line indicates the HOMO QP energy. (d) Normalized fragment polarization energy plotted as a function of the shell distance for the three chosen valence states of phenol.}
    \label{fig:fig2}
\end{figure}

First, we focus on the investigation of the H$_2$O environment. Figure~\ref{fig:fig2}a shows one of the localized orbitals on the H$_2$O molecules at $\sim$5 {\AA}. This is a hybridized state distributed on five molecules that are spatially separated. The total charge density for this H$_2$O shell (Figure~\ref{fig:fig2}b) then enters the $G_0W_0$ calculation and leads to the fragment correlation self-energy $\Sigma^k_{\rm{C}}$. As $\Sigma^k_{\rm{C}}$ is computed for multiple solvent molecules, we divide it by the number of solvent molecules in the shell and obtain a normalized fragment correlation self-energy $\overline{\Sigma}^k_{\rm{C}}$. The real part of $\overline{\Sigma}^k_{\rm{C}}$ for each H$_2$O shell is presented in Figure~\ref{fig:fig2}c. The labels from 1 to 6 correspond to the actual distances from 4 to 9 {\AA}. Around the peak of $\overline{\Sigma}^k_{\rm{C}}$ (the plasmon pole), the magnitude decays as the solvent molecules get further away. The vertical dashed line indicates the frequency where $\omega=\varepsilon_j$, i.e., the HOMO QP energy; the vertical coordinate of each intersection represents the actual polarization contribution from each shell to the QP energy. This derived polarization energy is denoted as $\overline{E}^k_{\rm{pol}}$. In Figure~\ref{fig:fig2}d, the solid red squares are the plot of the $\overline{E}^k_{\rm{pol}}$ induced by the HOMO (the 18$^{\rm{th}}$ valence state of phenol) at each distance, and the red line is fitted to this data set. The $\overline{E}^k_{\rm{pol}}$ decreases with increasing $\bR^k$ and practically vanishes at $\sim$9 {\AA}. 

Surprisingly, this is a very robust decaying pattern, as we observe a similar behavior from the first or the ninth valence state of phenol (half-filled squares with fitted lines in Figure~\ref{fig:fig2}d). All three $\overline{E}^{k}_{\rm{pol}}$ vanish at $\sim$9 {\AA}. The decaying pattern shows little dependence on the state's QP energy or its spatial distribution (Figure~\ref{fig:h2o_states}) on the solute molecule. We surmise that this pattern also applies to other small molecules similar to phenol, and further explanations are provided after discussing the nonaqueous solvents.

\begin{figure}
    \centering
    \includegraphics[width=\textwidth]{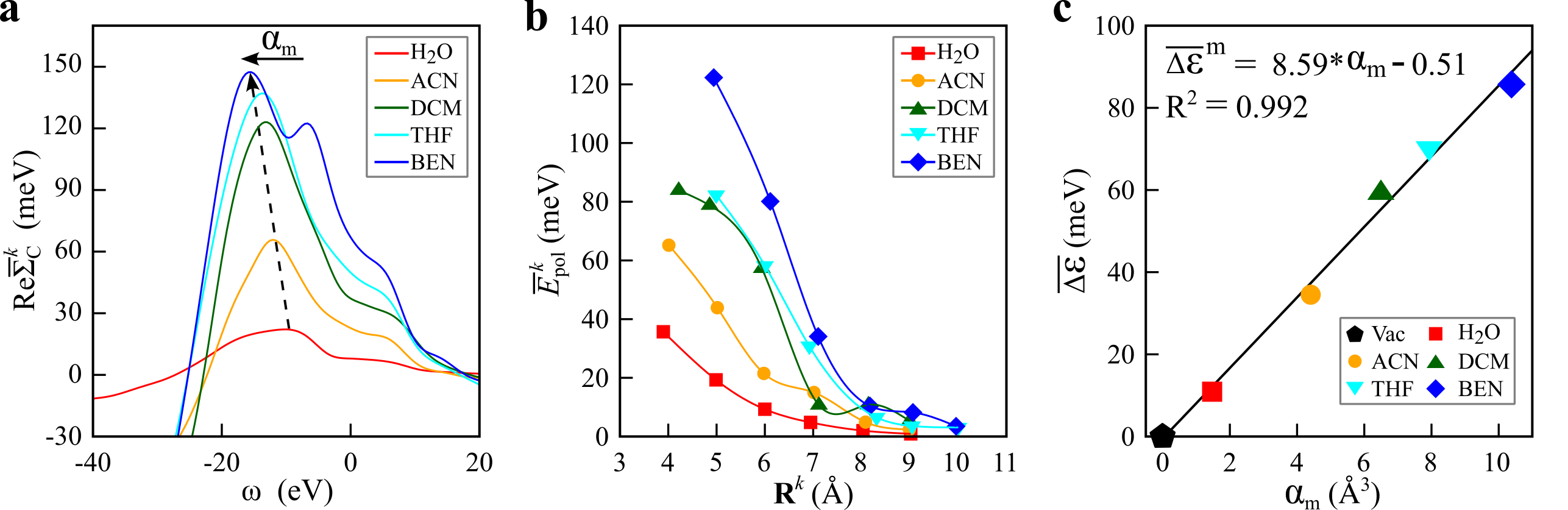}
    \caption{(a) Real part of the normalized fragment correlation self-energy for the shell at $\sim$5 {\AA} of each solvent. The dashed arrow indicates the shift of the pole, and the solid arrow denotes the order of $\alpha_m$. (b) Normalized fragment polarization energy plotted as a function of the shell distance for the HOMO state of phenol in each solvent. (c) Averaged IP shifts plotted as a function of the mean polarizability volume. The straight line is fitted using the linear regression model, where the slope is 8.59 meV/{\AA}$^3$ with an intercept of -0.51 meV at $\alpha_m = 0$.}
    \label{fig:fig3}
\end{figure}

The same fragmentation analysis is performed on the HOMO of the other four solvated systems (Figure~\ref{fig:shell_all}), in which the decays of $\overline{\Sigma}^k_{\rm{C}}$ of each individual solvent with respect to $\bR^k$ are consistent (Figure~\ref{fig:shell_SE2}). Further, we compare the $\overline{\Sigma}^k_{\rm{C}}$ from the shell at the same distance ($\sim$5 {\AA}) of each solvent, as  shown in Figure~\ref{fig:fig3}a: around the solvent plasmon pole (i.e., a polarizability resonance), the magnitude of $\overline{\Sigma}^k_{\rm{C}}$ has its maximum. At the same time, the pole shifts gradually to a higher frequency (indicated by the dashed arrow line), which corresponds to a faster response sustained by a more polarizable solvent. Further, the overall magnitude of $\overline{\Sigma}^k_{\rm{C}}$ increases with the polarizability volume of the liquid environment. In Figure~\ref{fig:fig3}b, the normalized polarization energy $\overline{E}^{k}_{\rm{pol}}$ is plotted with respect to $\bR^k$ for each solvent. Within 7 {\AA}, the magnitude of $\overline{E}^{k}_{\rm{pol}}$ along the vertical axis represents the strength of the interactions, which follows the order of $\alpha_m$ in Figure~\ref{fig:fig1}c. However, the decay of $\overline{E}^k_{\rm{pol}}$ along the horizontal axis does not differ too much among various solvents: particularly, the 9 {\AA} distance appears to be the common point at which the $\overline{E}^k_{\rm{pol}}$ for all the five solvents vanishes ($\overline{E}^k_{\rm{pol}}<10$ meV).

From the observations above, we find that the macroscopic solvent polarizability dictates the temporal behavior (i.e., how fast the response is) and the strength of the response (i.e., the magnitude of the nonlocal correlation self-energy). A common cutoff distance, within which the solvent molecules are considered \textit{effective}, is found at $\sim$9 {\AA} away from the phenol solute. This effective interacting radius depends neither on the ionization state nor the solvent type, and this phenomenon, as we believe, can be attributed to the localized molecular excitation on phenol. In practice, electronic states on the solvent molecules are only mildly perturbed by the excitation of the solute. For solvent molecules at long distances, the perturbation is even weaker. Hence, the 9 {\AA} cutoff should be a consequence of the localized molecular excitation and applies likely to other molecules of similar size to phenol.

Based on the effective interacting radius, we identify the number of \textit{effective solvent molecules} $N_{\rm{eff}}$ for each solvated system (see Table~\ref{tab:num_eff_solv}). From H$_2$O to BEN (as ordered in Figure~\ref{fig:fig1}c), the $N_{\rm{eff}}$ (averaged over five snapshots) are found to be 97, 32, 24, 19, and 16. This will help to explain the apparent contradiction observed for the DCM solvent (Figure~\ref{fig:fig1}c): although DCM is less polarizable than THF and BEN, it leads to stronger energy shifts because of a larger $N_{\rm{eff}}$ (i.e., more effective solvent molecules). The difference in $N_{\rm{eff}}$ originates from the molecular size and mass density difference among various solvents. Moreover, we divide the QP energy shifts $\Delta \varepsilon$ by the number of effective solvent molecules $N_{\rm{eff}}$ and denote the result as $\xoverline[0.8]{\Delta \varepsilon}$. The $\xoverline[0.8]{\Delta \varepsilon}$ represents the IP shift per effective solvent molecule. In Figure~\ref{fig:fig3}c, the derived $\xoverline[0.8]{\Delta \varepsilon}$ is plotted as a function of $\alpha_m$, with $\alpha_m = \xoverline[0.8]{\Delta \varepsilon} = 0$ representing the solvent-free (i.e., solute placed in a vacuum) case. This data set is fitted to the linear regression model, resulting in a determination $R^2>0.99$. The $\xoverline[0.8]{\Delta \varepsilon}$ is shown to depend linearly on $\alpha_m$ with a slope of 8.59 meV/\AA$^3$ and an interception of $-0.51$ meV at $\alpha_m = 0$.

The linear relationship between $\xoverline[0.8]{\Delta \varepsilon}$ and $\alpha_m$ shown in Figure~\ref{fig:fig3}c can lead to a potential solvation model for computing the QP excitation energies, corresponding to the ionization potentials in this particular case, of molecules in an arbitrary solvent environment. The current work presents a model parameterization for phenol and its generalization to different solutes is discussed below
\begin{equation}
\begin{split}
  \varepsilon^{\rm{solv}} &=\varepsilon^{\rm{iso}}+\Delta \varepsilon^{m} \\
  &= \varepsilon^{\rm{iso}} + N_{\rm{eff}} {\xoverline[0.8]{\Delta \varepsilon}}^m
\end{split}
\label{eq:solv_model}
\end{equation}
where $\varepsilon^{\rm{iso}}$ denotes the QP energy of the isolated molecule from first-principles calculations, and $\Delta \varepsilon^{m}$, derived as a product of $N_{\rm{eff}}$ and ${\xoverline[0.8]{\Delta \varepsilon}}^m$, is the QP energy shifts (i.e., the IP shifts) induced by the solvent environment. Using the 9 {\AA} cutoff radius, the number of effective solvent molecules $N_{\rm{eff}}$ can be determined by sampling snapshots from MD simulations (denoted $N_{\rm{eff}}^0$), or estimated by the solvent's mass density and the solute's exclusion volume\cite{excl_vol} (denoted $N_{\rm{eff}}^m$). The ${\xoverline[0.8]{\Delta \varepsilon}}^m$ in eq~\eqref{eq:solv_model} is obtained by inserting the solvent polarizability volume $\alpha_m$ (derived from $n_r$) into the linear equation shown in Figure~\ref{fig:fig3}c. The $\Delta \varepsilon^m$ in eq~\eqref{eq:solv_model} can thus be derived from only classical solvent properties, which are accessible for most of the common solvents. Although we hypothesize that the linear equation shown in Figure~\ref{fig:fig3}c should be universal, the exact proportionality coefficient would likely change for another solute molecule since the solute-solvent interactions are coupled. For instance, a more polarizable solute might cause stronger response from the solvent environment, leading to a larger coefficient than the one found in Figure~\ref{fig:fig3}c.

We use the proposed solvation model to compute the IP shifts for phenol in the five investigated solvents and compare them with the first-principles results. The $\alpha_m$ listed in Figure~\ref{fig:fig1}c are inserted into the linear equation in Figure~3c to derive ${\xoverline[0.8]{\Delta \varepsilon}}^m$, which show differences from the first-principles values (Figure~\ref{fig:fit_check}). For the number of effective solvent molecules, we consider both $N_{\rm{eff}}^m$ and $N_{\rm{eff}}^0$ (defined above). Numerical details are provided in Table~\ref{tab:solv_model_test}. The computed results using the combination of $N_{\rm{eff}}^m$ and ${\xoverline[0.8]{\Delta \varepsilon}}^m$ are plotted in Figure~\ref{fig:fig1}c ($\Delta \varepsilon^m_1$, red squares). For the first four solvents, the derived trend agrees well with the first-principles one (blue circles), despite a common overestimation of $\sim$0.1 eV. The deviation of $\Delta \varepsilon^m_1$ for BEN is more significant due to an overestimated $N_{\rm{eff}}^m$. By replacing $N_{\rm{eff}}^m$ with $N_{\rm{eff}}^0$ ($\Delta \varepsilon^m_2$, orange diamonds), the derived IP shifts for the BEN solvent become closer to the first-principles result. For the DCM solvent, $\Delta \varepsilon^m_2$ underestimates the IP shifts due to a smaller ${\xoverline[0.8]{\Delta \varepsilon}}^m$ than the first-principles ${\xoverline[0.8]{\Delta \varepsilon}}^0$ (Figure~\ref{fig:fit_check}). However, the agreement with the first-principles results is generally improved when the $N_{\rm{eff}}$ is estimated by MD simulations.

For further development and optimization of this simple solvation model, we note that (1) the full parameterization relies on exploring multiple solute molecules to elucidate the solute-dependence discussed above; (2) the linear equation in Figure~\ref{fig:fig3}c can be improved, e.g., by sampling more solvent cases and averaging over more MD snapshots.

In summary, this work introduces a generalized decomposition scheme of the $GW$ correlation self-energy based on the fragmentation of a multi-molecule system. This methodology is employed to investigate the energy shifts of single-(quasi)particle excitations in various solvent environments. The fragmentation (decomposition) of the correlation self-energy is formulated upon the assumption that the induced density fluctuations, represented by Pipek-Mezey localized orbitals, are local on each fragment. 

We sample molecular dynamics simulations using water and four other organic solvents together with the phenol solute. The first observation is that the explicit IP shifts do not follow the order of the mean polarizability volume computed for each solvent. To investigate this disagreement, we apply the self-energy fragmentation scheme to explore the solvation many-body effects in these systems. Specifically, we compute the fragment correlation contributions for solvation shells at various distances away from the solute molecule. The fragment correlation self-energy decays monotonically as a function of the shell distance. At the distance of $\sim$9 {\AA}, the correlation contribution practically vanishes. This distance corresponds to an effective interacting radius for considering the solvent response stemming from the induced dipole interactions. The 9 {\AA} effective interacting radius depends neither on the ionization state of the solute nor on the solvent type. This phenomenon can be attributed to the localized feature of QP excitations on the solute. 

Comparing the correlation self-energies among various environments, we find that the macroscopic solvent polarizability is directly related to the temporal behavior and the correlation strength. However, the apparent 9 {\AA} cutoff radius is unaffected by the polarizability of the solvent. This existing cutoff radius indicates that the explicit solvation many-body effects on the ionization energy depend not only on the polarizability but also on the number of effective solvent molecules. The latter can be extremely sensitive to practical experimental conditions. Indeed, if both the polarizability and the number of solvent molecules within the effective volume are considered together, the average IP shift is perfectly reflected by the mean polarizability volume and exhibits a linear dependence. Based on this connection, we further propose a possible solvation model to compute the QP energies of solvated molecules. This model requires only classical parameters that are readily accessible.

Although further explorations are needed to confirm the universality of the effective interacting radius and the solvation model, rich information about the solvation many-body effects has been unveiled by the proposed self-energy fragmentation method. We believe this approach will provide a powerful tool for understanding and analyzing the interactions and couplings between specific fragments in composite condensed systems.

\begin{acknowledgement}
This work was supported by the NSF CAREER award through Grant No. DMR-1945098. The calculations were performed as part of the XSEDE computational Project No. TG-CHE180051. Use was made of computational facilities purchased with funds from the National Science Foundation (CNS-1725797) and administered by the Center for Scientific Computing (CSC). The CSC is supported by the California NanoSystems Institute and the Materials Research Science and Engineering Center (MRSEC; NSF DMR 1720256) at UC Santa Barbara.
\end{acknowledgement}

\begin{suppinfo}
Texts: details of computations and methodologies; Tables: parameters in the DFT and $GW$ calculations, number of solvent molecules in the small and large simulation cells, QP energies and IP shifts of each snapshot in each solvated system, number of effective solvent molecules in each snapshot of each solvated system; Figures: extracted snapshots from the MD trajectories, fluctuations of the QP energies and IP shifts with respect to the MD snapshot, IP shifts as a function of the mean polarizability, solvation shells at various distances, fragment correlation self-energy of different ionization states and solvents, graphical solutions to the QP energy of the isolated and solvated phenol.
\end{suppinfo}

\bibliography{solvent}

\begin{mcitethebibliography}{88}
\bibitem[exc()]{excl_vol}
The solute exclusion volume by phenol is calculated as the volume of a sphere
  with a 3.5 {\AA} radius.\relax
\mciteBstWouldAddEndPunctfalse
\mciteSetBstMidEndSepPunct{\mcitedefaultmidpunct}
{}{\mcitedefaultseppunct}\relax
\EndOfBibitem
\end{mcitethebibliography}

\end{document}